\documentclass[11pt,twoside]{article}
\usepackage{asp2010}

\resetcounters

\begin{document}

\title{The Fundamental Plane of Galaxy Group Mergers}
\author{Dan Taranu, John Dubinski, and Howard Yee
\affil{Department of Astronomy and Astrophysics, University of Toronto}}

\begin{abstract}
We present a series of hundreds of collisionless simulations of galaxy group mergers. These simulations are designed to test whether the properties of elliptical galaxies - including the key fundamental plane scaling relation, morphology and kinematics - can be simultaneously reproduced by dry multiple mergers in galaxy groups. Preliminary results indicate that galaxy group mergers can produce elliptical remnants lying on a tilted fundamental plane, even without a central dissipational component from a starburst. This suggests that multiple mergers in groups are an alternate avenue for the formation of elliptical galaxies which could well dominate for luminous ellipticals.
\end{abstract}

\section{Introduction}
We aim to test the hypothesis that ellipticals form through hierarchical mergers of spiral galaxies in groups. If so, is the observed 'tilt' of the fundamental plane from the virial relation a natural consequence of hierarchical merging of spirals in $\Lambda$CDM cosmology? Does the scaling relation of spirals - the Tully-Fisher relation - imprint itself in elliptical properties as well? If so, this would contrast with evidence presented by e.g. \cite{RobCoxHer06}, who suggest that elliptical properties (including the tilt of the fundamental plane) arise from varying gas fractions in binary mergers of spirals.

Our method is to create a large statistical sample of galaxy merger simulations. These simulations naturally include mechanisms investigated independently in previous works, including minor merging, hierarchical merging and group mergers. Similar simulations have been performed by \cite{AceVel05}. The chief improvement in this work is to use equilibrium, bar-stable galaxy models based on M31 data \citep{WidDub05,WidPymDub08} and to fairly compare the results of hundreds of simulations to observations of local ellipticals (e.g. SDSS and Atlas3D).

\section{Simulations}
The initial conditions are designed to be quasi-cosmological - not an unbiased sample but a roughly even sampling of the parameter space likely to produce groups with central ellipticals. Briefly, we create simulations using galaxy models with Sersic $n_{s}=1$ 'pseudobulges' and $n_{s}=4$ classical/de Vaucouleurs bulges. Each model has 2 sets of simulation with 7 target luminosity bins from 1/8 to 8 L* (in factors of 2), with 8 groups in each mass bins for a total of 2x7x8=108 simulations. Galaxies are placed within a sphere of $r=r_{200}$ of the group at z=1 (or twice $r_{200}$ at z=1) with the most massive galaxy in the center. Each group has between $N_{gals} \ge N_{min}=3$ and $N_{gals} \le N_{max}=10 \cdot sqrt(L/L*)$. Within each group luminosity bin, the number of galaxies in each simulation varies linearly from the minimum (3) to the maximum, so that L* groups have between 3 and 10 galaxies and the largest groups have 28 galaxies. Galaxy luminosities are randomly sampled from an observed spiral luminosity (Schechter) function. Luminosities are drawn from a restricted range of the luminosity function with a width equal to $(N_{gals}+2)/10$ dex and such that the integral under the curve is equal to the target group luminosity. This setup excludes groups with a dominant bright galaxy and much smaller satellites (like the Milky Way and Magellanic clouds) which are unlikely to form ellipticals. A control sample of equal mass mergers is also created in each mass bin, one group using 3 galaxies and the other $N_{max}$. Satellites are given preferentially inward and radial orbits, with the group as a whole being subvirial (to ensure collapse) and no satellite having $|v| > v_{escape}$.

\begin{figure*}
\includegraphics[width=0.3\textwidth]{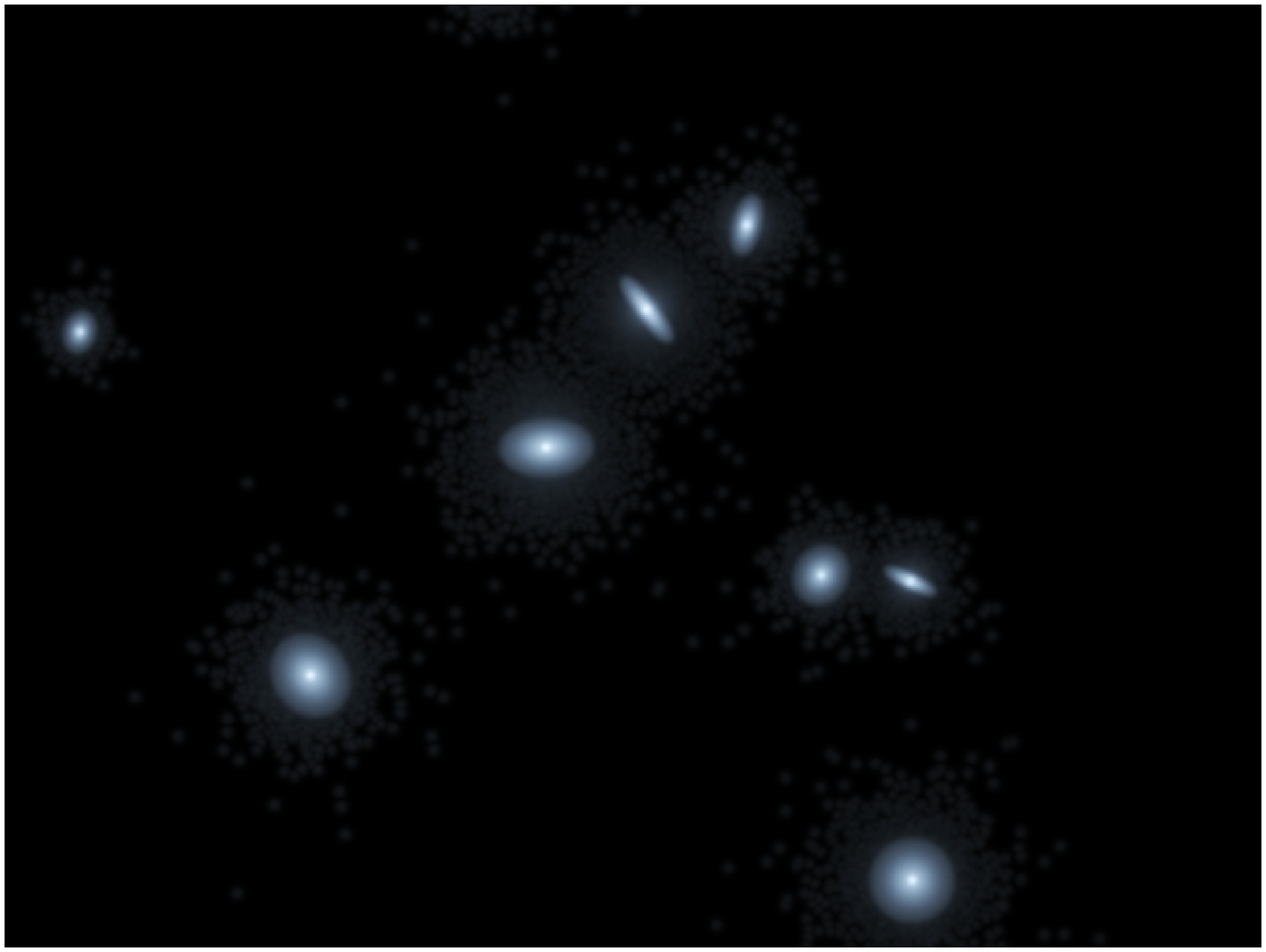}
\includegraphics[width=0.3\textwidth]{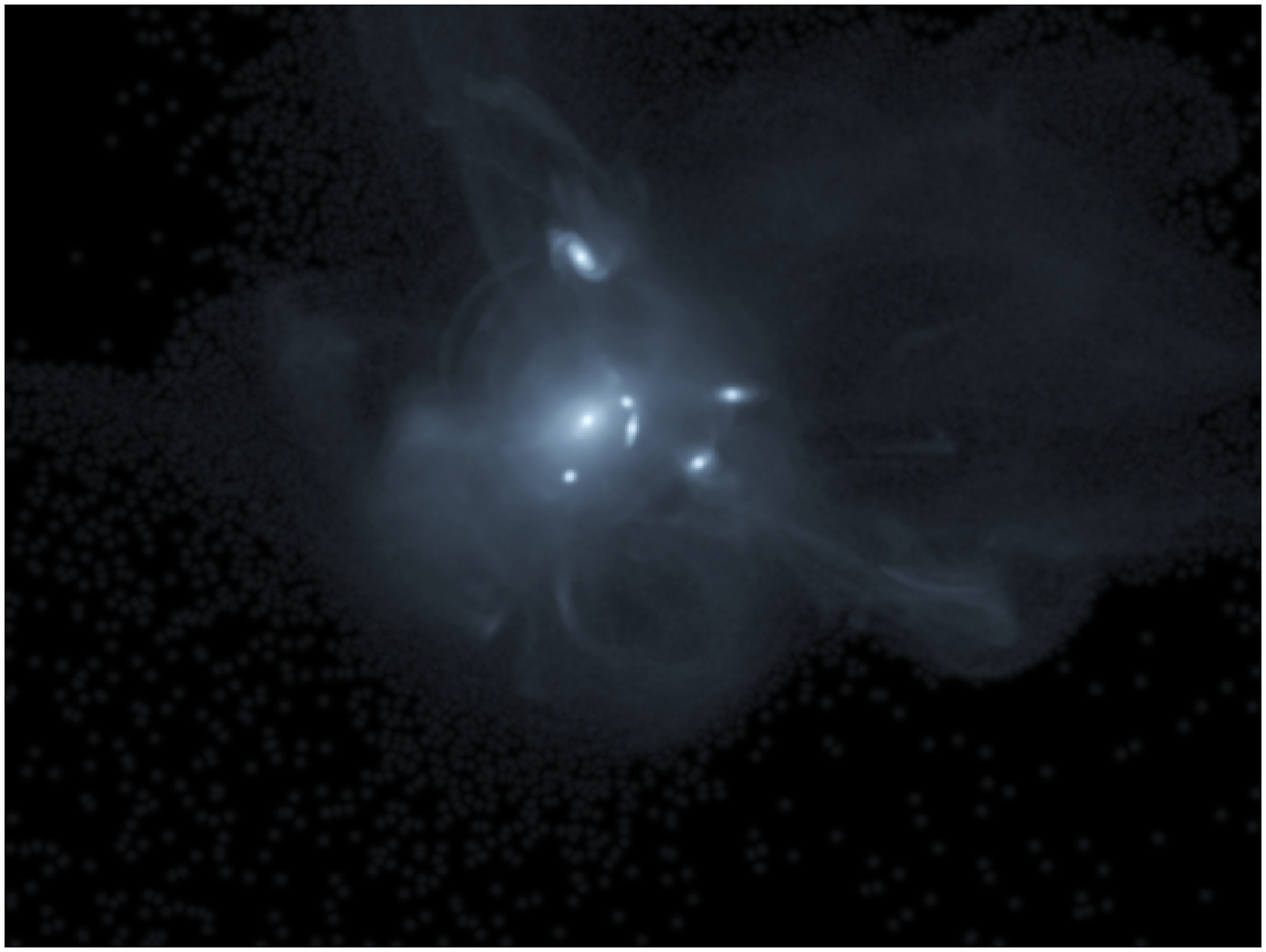}
\includegraphics[width=0.3\textwidth]{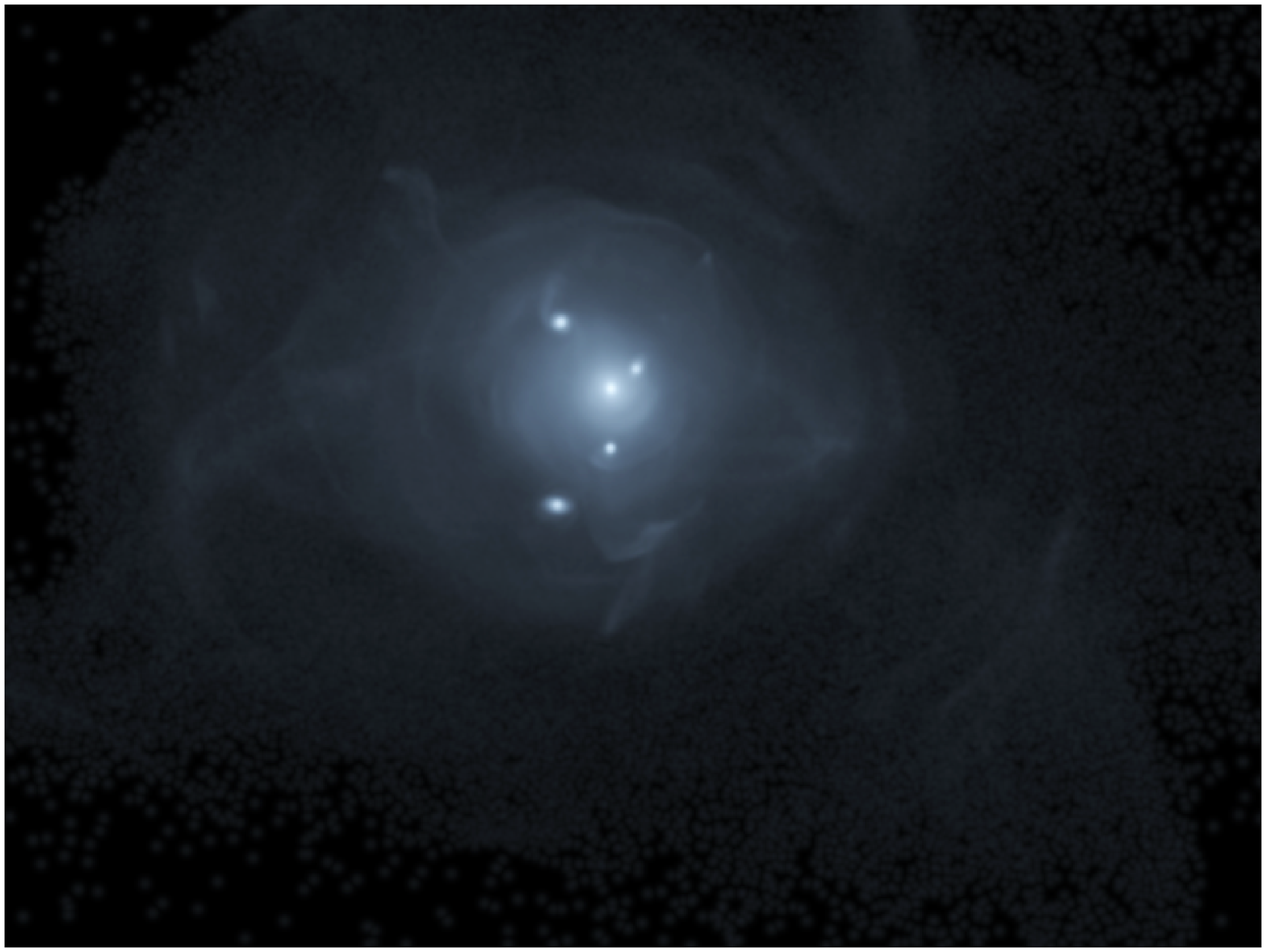}
\caption{Images of a high-resolution re-simulation of a group of 10 galaxies totaling L*, starting from initial conditions and advancing 25,000 timesteps (~5 Gyr) in each frame.} 
\label{fig:images}
\end{figure*}

Groups are simulated for 10 Gyr with the parallel tree code Partree \citep{Dub96}, using 50,000 timesteps and a softening length (spatial resolution) of 100 pc. Figure \ref{fig:images} shows a typical evolution for one such group. Simulations are analyzed after 5, 7.5 and 10 Gyr. For brevity, only results for the 10 Gyr evolution (assuming groups formed at z=2) are shown. We note that groups with fewer galaxies typically merge in a few Gyr, while groups with more galaxies continue (slowly) accreting lower-mass satellites up to and beyond 10 Gyr.
We process simulations with our own imaging pipeline, which is intended to create SDSS-like images of our groups, using sky backgrounds and S/N typical for SDSS imaging runs of galaxies at z=0.025. We also create kinematical maps at the same spatial resolution but with no noise in the velocity space. Each galaxy is imaged in ten equally spaced projections to boost our sample size by a factor of ten.

\section{Results}

\begin{figure*}
\includegraphics[width=0.50\textwidth]{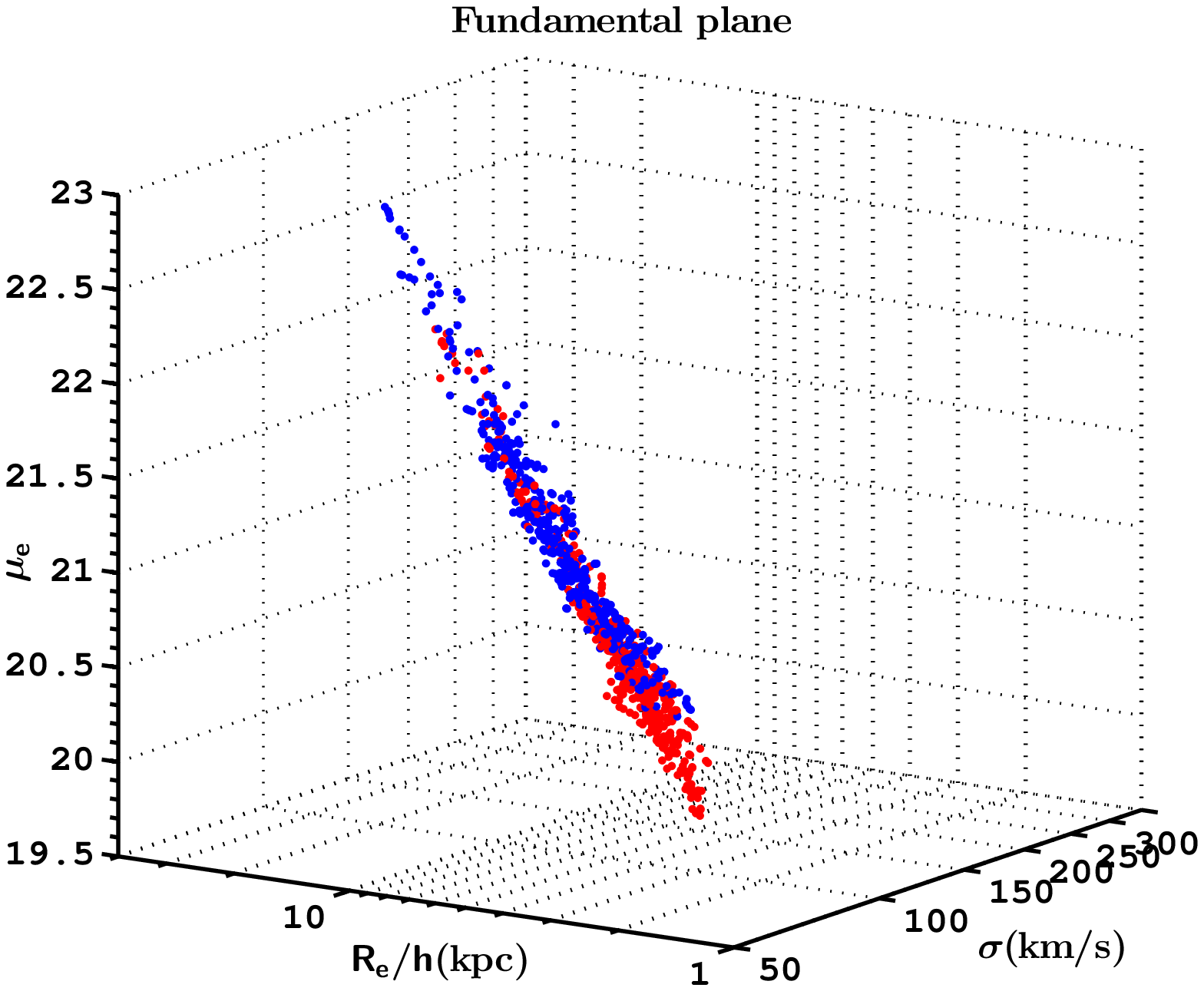}
\includegraphics[width=0.50\textwidth]{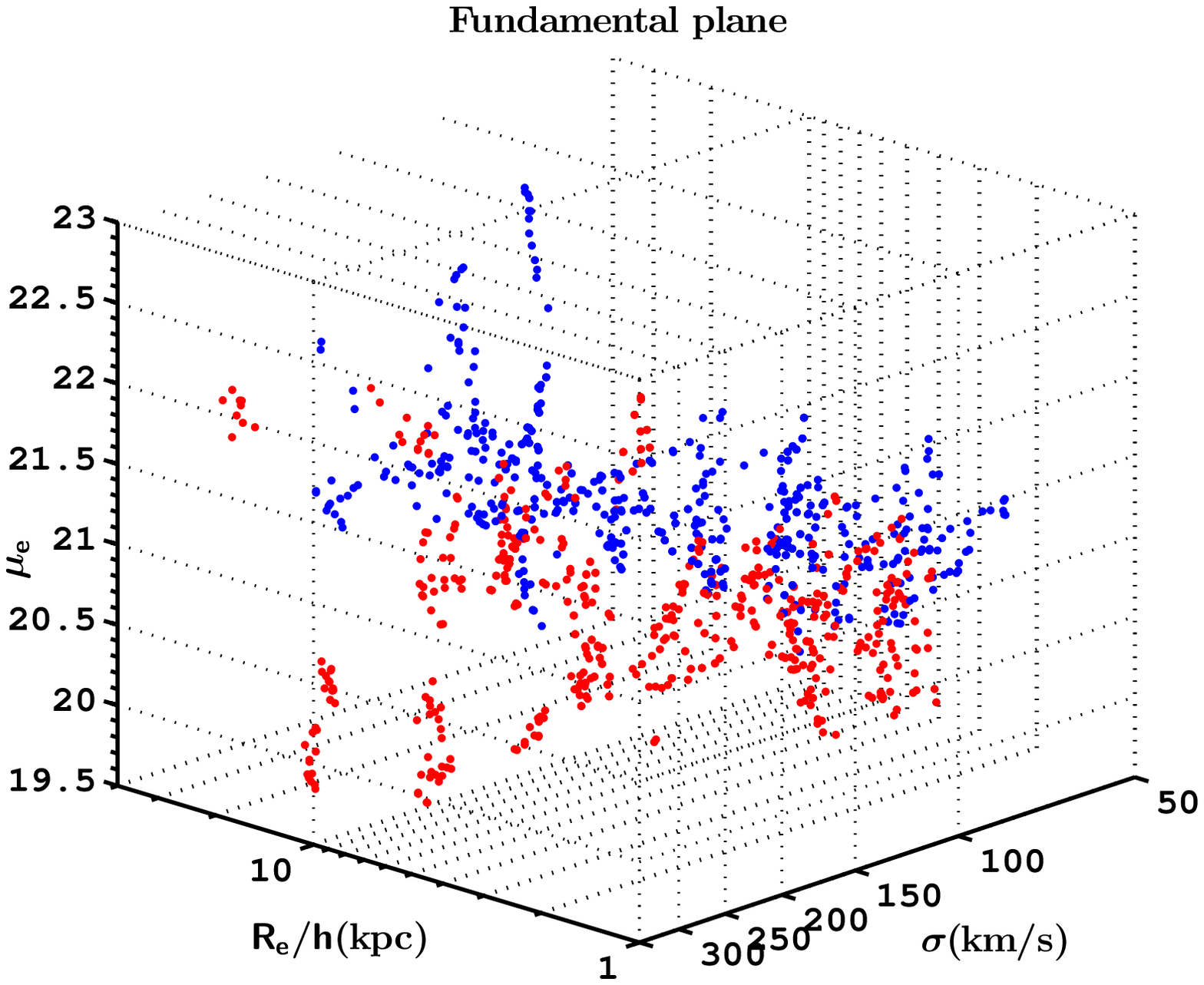}
\caption{Face-on and edge-on projections of the fundamental plane of classical bulge mergers.} 
\label{fig:fplane}
\end{figure*}

Our main result is a measurement of the fundamental plane relation: $log(R_{eff}) = a\log(\sigma) + b\mu + c$. The tilt for our various samples is tabulated in table~\ref{tab:fplanet}. Both pseudobulge and classical bulge mergers create a tilted fundamental plane. Figure~\ref{fig:fplane} illustrates the small scatter and spatial extent of the plane. This challenges the interpretation of \cite{HopCoxHer08} that dissipation is necessary for the tilt, even if it is sufficient. Interestingly, equal mass mergers appear create a tilt. This could indicate that the tilt is a generic feature of multiple merging rather than being dependent on a particular spiral scaling relation; however, we caution that the equal-mass merger sample is quite small. 

\begin{table*} [h]
\caption{Best fit fundamental plane coefficients of central galaxies in the simulations compared to SDSS observations \citep{HydBer09}.}
\label{tab:fplanet}

\begin{tabular}{cccccc}
\hline
Sample & N & a & b & observed rms & intrinsic rms\\
\hline
Virial & n/a & 2 & 0.4 & n/a & n/a \\
Classical bulge & 101 & 1.65 & 0.296 & 0.040 & n/a\\
Exponential bulge & 85 & 1.73 & 0.327 & 0.052 & n/a\\
Classical bulge, equal mass galaxies & 33 & 1.74 & 0.291 & 0.046 & n/a\\
Exponential bulge, equal mass galaxies & 29 & 1.72 & 0.295 & 0.054 & n/a\\
SDSS & ~50,000 & 1.43 & 0.315 & 0.066 & 0.058\\
SDSS, stellar mass (excluding M$_{*}$/L) & ~50,000 & 1.63 & 0.336 & 0.065 & 0.049\\
\hline
\end{tabular}
\end{table*}

The tilt appears similar to that from SDSS observations \citep{HydBer09} excluding the M$_{*}$/L contribution, which is nil by construction in our simulations.

All galaxies in each image are fit with a single Sersic profile, although only central ellipticals are included in the final catalog. Classical bulges have higher central densities than exponential bulges, so we would expect the inner bulge-dominated profile of classical bulge remnants to be steeper with larger $n_{s}$. Equivalently, exponential bulges should be unable to form steep inner profiles with large $n_{s}$ due to the conservation of phase space density \citep{Car86}.

\begin{figure*}
\includegraphics[width=0.50\textwidth]{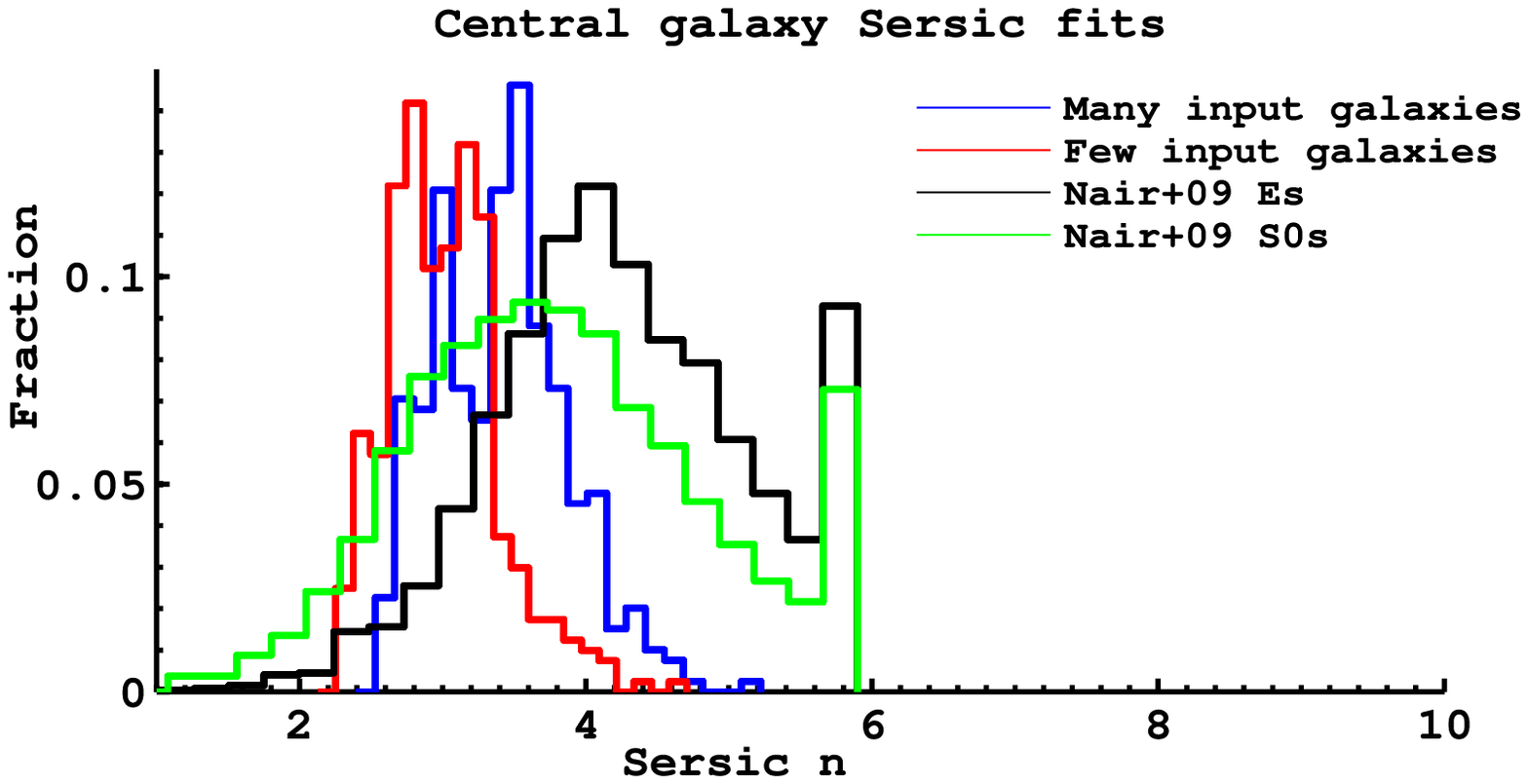}
\includegraphics[width=0.50\textwidth]{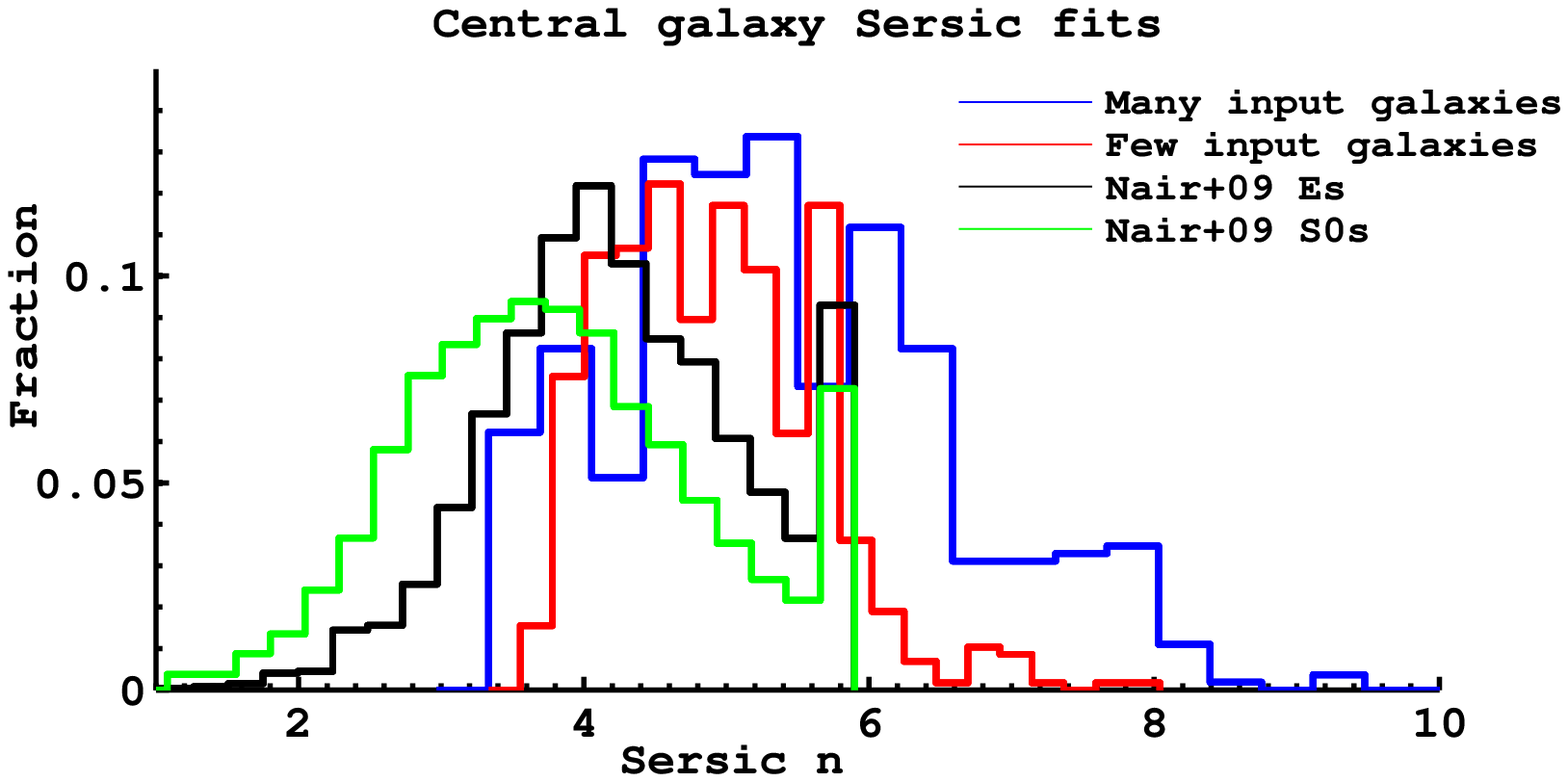}
\includegraphics[width=0.50\textwidth]{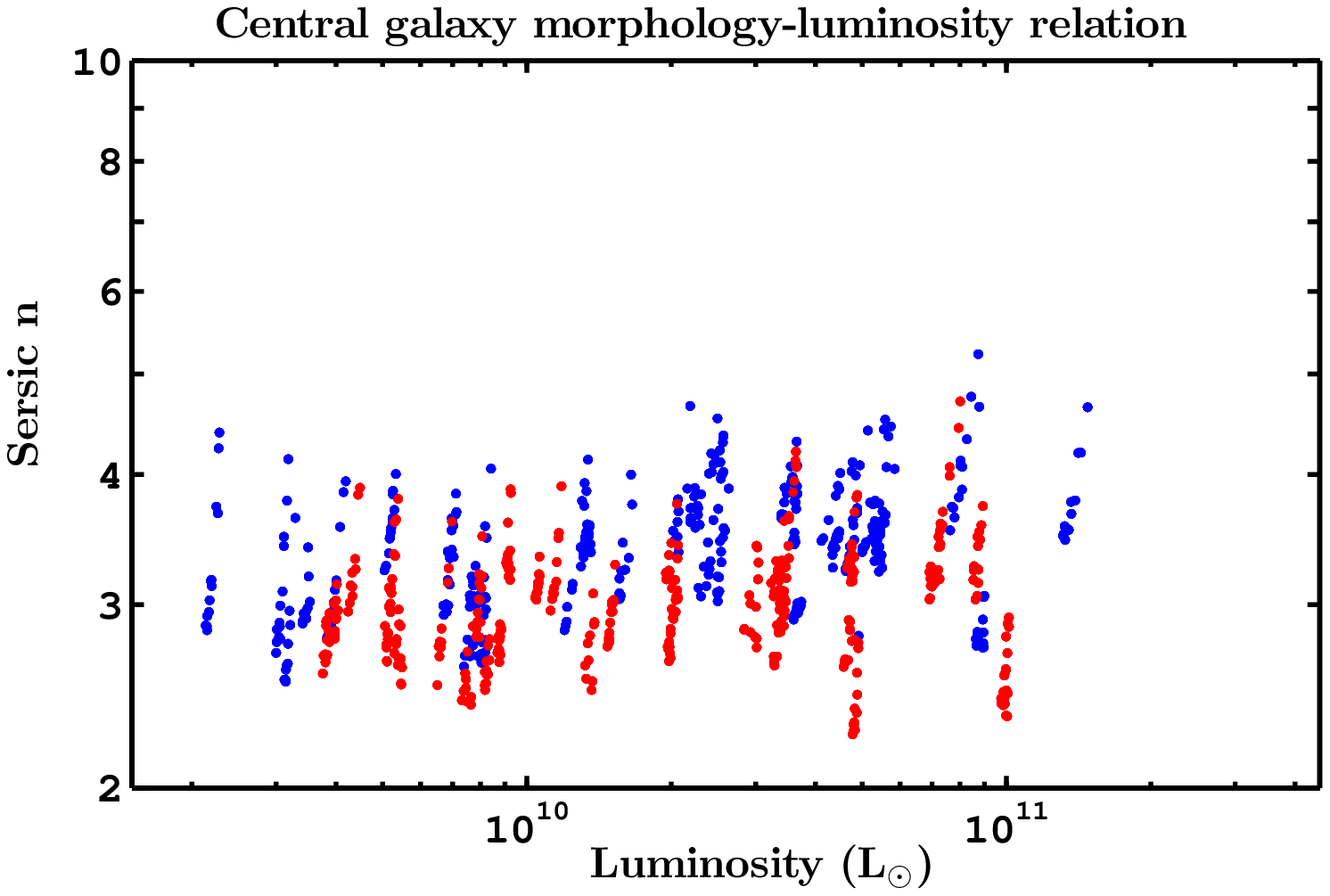}
\includegraphics[width=0.50\textwidth]{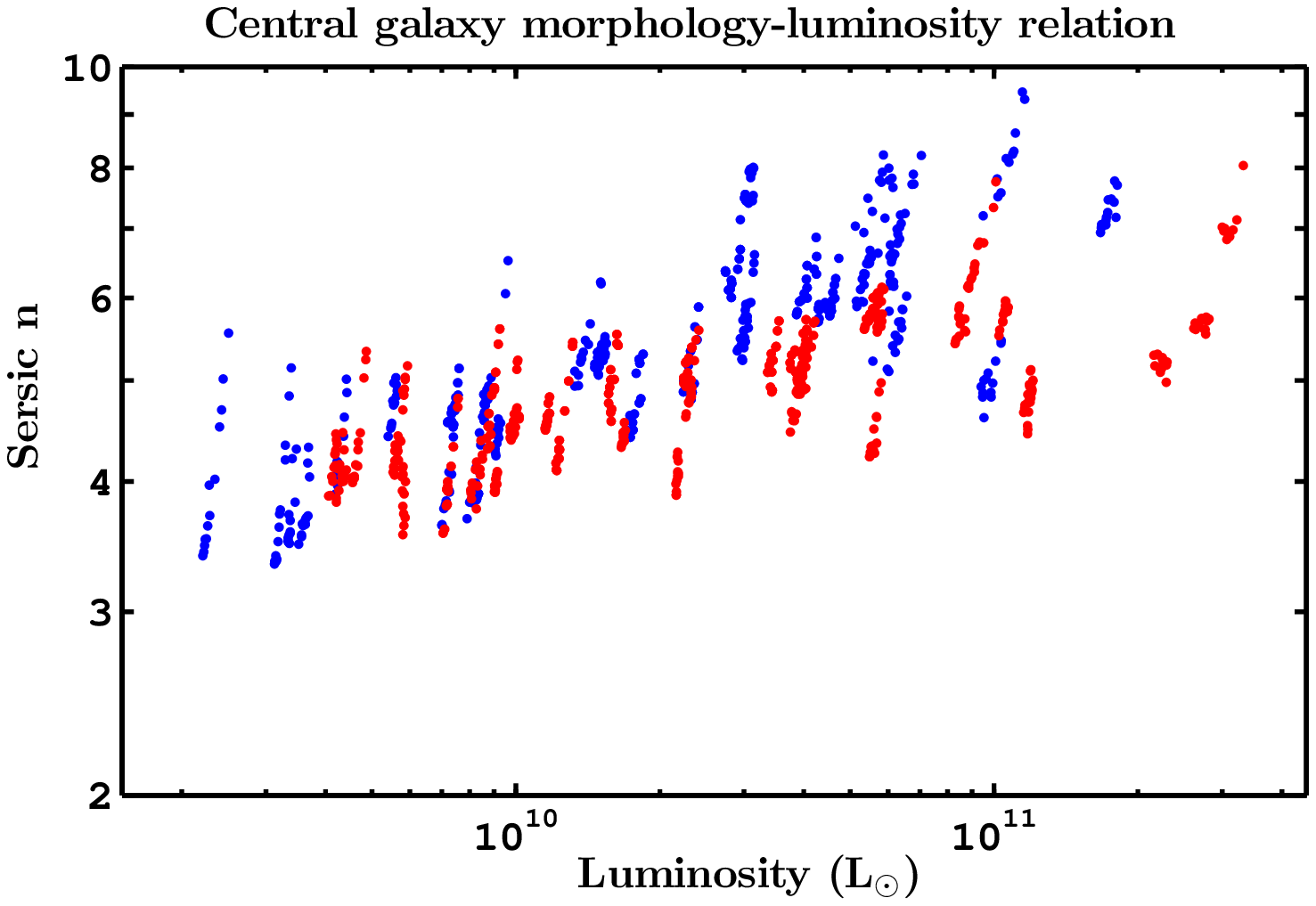}
\caption{Sersic indices of central ellipticals. Classical bulge mergers (right) have larger $n_{s}$ for the same initial conditions and show a strong dependence of $n_{s}$ on galaxy luminosity, as with observed ellipticals but unlike exponential bulge mergers (left). Observed data from \cite{NaiAbr10}.} 
\label{fig:sersicn}
\end{figure*}

We confirm that classical bulge mergers have larger $n_{s}$ and a distribution more consistent with observations than exponential bulge mergers (Fig.~\ref{fig:sersicn}). Classical bulge mergers appear to have slightly larger $n_{s}$ than the observations and exponential bulge mergers considerably lower. This suggests that ellipticals - if they are formed by dry mergers - are formed from spirals with a range of bulge profiles, consistent with measurements of spiral bulge profiles.

Only classical group mergers produce a correlation between luminosity and $n_{s}$. This relation is a result of more luminous ellipticals having been produced on average by more mergers. The dependence of merger rate on halo mass is a prediction of $\Lambda$CDM (e.g. \cite{HopCroBun10}). However, exponential bulges are simply not concentrated enough to create merger remnants with $n_{s} > 4$, even with repeated merging. Thus, luminous ellipticals are unlikely to be the product of exponential bulge mergers.

We measure the traditional $v/\sigma$ measure as the luminosity-weighted average within $R_{eff}$, as used in IFU observations like Atlas3D \citep{CapEmsKra11}. As in Atlas3D, most of our ellipticals are slow rotators  However, some remnants are formed with $v/\sigma$ as large as 0.35, showing that it is possible to form fast rotators from dry mergers alone - a contrast with previous studies of dry binary mergers, which only formed very slow rotators.

There does not appear to be any strong correlation between rotational support and number of mergers, galaxy luminosity or other parameters. We suspect the rotation is due to net angular momentum in the group. Whatever the case, it is possible for multiple mergers to produce fast rotators. However, fast-rotating S0s have been observed with $v/\sigma > 0.5$. If these S0s are merger products, they cannot have been formed from dry mergers alone.

\bibliography{author}

\begin{thebibliography}{}
\expandafter\ifx\csname natexlab\endcsname\relax\def\natexlab#1{#1}\fi
\expandafter\ifx\csname url\endcsname\relax
  \def\url#1{\texttt{#1}}\fi
\expandafter\ifx\csname urlprefix\endcsname\relax\def\urlprefix{URL }\fi
\providecommand{\eprint}[2][]{\url{#2}}

\bibitem[{{Aceves} \& {Vel{\'a}zquez}(2005)}]{AceVel05}
{Aceves}, H., \& {Vel{\'a}zquez}, H. 2005, \mnras, 360, L50.
  \eprint{arXiv:astro-ph/0412307}

\bibitem[{{Cappellari} et~al.(2011){Cappellari}, {Emsellem}, {Krajnovi{\'c}},
  {McDermid}, {Scott}, {Verdoes Kleijn}, {Young}, {Alatalo}, {Bacon}, {Blitz},
  {Bois}, {Bournaud}, {Bureau}, {Davies}, {Davis}, {de Zeeuw}, {Duc},
  {Khochfar}, {Kuntschner}, {Lablanche}, {Morganti}, {Naab}, {Oosterloo},
  {Sarzi}, {Serra}, \& {Weijmans}}]{CapEmsKra11}
{Cappellari}, M., {Emsellem}, E., {Krajnovi{\'c}}, D., {McDermid}, R.~M.,
  {Scott}, N., {Verdoes Kleijn}, G.~A., {Young}, L.~M., {Alatalo}, K., {Bacon},
  R., {Blitz}, L., {Bois}, M., {Bournaud}, F., {Bureau}, M., {Davies}, R.~L.,
  {Davis}, T.~A., {de Zeeuw}, P.~T., {Duc}, P.-A., {Khochfar}, S.,
  {Kuntschner}, H., {Lablanche}, P.-Y., {Morganti}, R., {Naab}, T.,
  {Oosterloo}, T., {Sarzi}, M., {Serra}, P., \& {Weijmans}, A.-M. 2011, \mnras,
  413, 813. \eprint{1012.1551}

\bibitem[{{Carlberg}(1986)}]{Car86}
{Carlberg}, R.~G. 1986, \apj, 310, 593

\bibitem[{{Dubinski}(1996)}]{Dub96}
{Dubinski}, J. 1996, New Astronomy, 1, 133. \eprint{arXiv:astro-ph/9603097}

\bibitem[{{Hopkins} et~al.(2008){Hopkins}, {Cox}, \& {Hernquist}}]{HopCoxHer08}
{Hopkins}, P.~F., {Cox}, T.~J., \& {Hernquist}, L. 2008, \apj, 689, 17.
  \eprint{0806.3974}

\bibitem[{{Hopkins} et~al.(2010){Hopkins}, {Croton}, {Bundy}, {Khochfar}, {van
  den Bosch}, {Somerville}, {Wetzel}, {Keres}, {Hernquist}, {Stewart},
  {Younger}, {Genel}, \& {Ma}}]{HopCroBun10}
{Hopkins}, P.~F., {Croton}, D., {Bundy}, K., {Khochfar}, S., {van den Bosch},
  F., {Somerville}, R.~S., {Wetzel}, A., {Keres}, D., {Hernquist}, L.,
  {Stewart}, K., {Younger}, J.~D., {Genel}, S., \& {Ma}, C.-P. 2010, \apj, 724,
  915. \eprint{1004.2708}

\bibitem[{{Hyde} \& {Bernardi}(2009)}]{HydBer09}
{Hyde}, J.~B., \& {Bernardi}, M. 2009, \mnras, 396, 1171. \eprint{0810.4924}

\bibitem[{{Nair} \& {Abraham}(2010)}]{NaiAbr10}
{Nair}, P.~B., \& {Abraham}, R.~G. 2010, \apjs, 186, 427. \eprint{1001.2401}

\bibitem[{{Robertson} et~al.(2006){Robertson}, {Cox}, {Hernquist}, {Franx},
  {Hopkins}, {Martini}, \& {Springel}}]{RobCoxHer06}
{Robertson}, B., {Cox}, T.~J., {Hernquist}, L., {Franx}, M., {Hopkins}, P.~F.,
  {Martini}, P., \& {Springel}, V. 2006, \apj, 641, 21.
  \eprint{arXiv:astro-ph/0511053}

\bibitem[{{Widrow} \& {Dubinski}(2005)}]{WidDub05}
{Widrow}, L.~M., \& {Dubinski}, J. 2005, \apj, 631, 838.
  \eprint{arXiv:astro-ph/0506177}

\bibitem[{{Widrow} et~al.(2008){Widrow}, {Pym}, \& {Dubinski}}]{WidPymDub08}
{Widrow}, L.~M., {Pym}, B., \& {Dubinski}, J. 2008, \apj, 679, 1239.
  \eprint{0801.3414}

\end{thebibliography}

\end{document}